\documentclass{aa}  

\usepackage{graphicx}
\usepackage{txfonts}
\usepackage{aalongtable,lscape}

\usepackage{natbib}
\bibpunct{(}{)}{;}{a}{}{,}
\begin{document}
  \title{Spectroscopy of globular clusters in the low-luminosity spiral galaxy NGC~45\thanks{
Based on data collected at the Cerro Paranal, Chile run by the  European Southern Observatory 
(ESO) under  programme ID 077.D-0403(A) and  077.D-0403(B).}}

%   \subtitle{Abundances}

\author {
  M.~D.~Mora\inst{1,3} \and
  S.~S.~Larsen\inst{2} \and
  M.~Kissler-Patig\inst{3} 
}

\authorrunning{M.~D.~Mora et al.}
\titlerunning{Spectroscopy of globular clusters in NGC~45.}
 \offprints{M.~D.~Mora}

 \institute
     {   
       Departamento de F\'isica y Astronom\'ia, Facultad de Ciencias, Universidad de Valpara\'iso. 	Av. Gran Breta\~na 1111, Valpara\'iso, Chile.  
       \\
       \email{mmora@dfa.uv.cl}
       \and
       Astronomical Institute, University of Utrecht, Princetonplein 5,
       NL-3584 CC, Utrecht, The Netherlands.
       \\
       \email{S.S.Larsen@uu.nl}
       \and
     European Southern Observatory, Karl-Schwarzschild-Strasse 2, 85748 Garching bei M\"unchen, Germany.
     \\
     \email{mkissler@eso.org}
     }
     
     \date{Received  / Accepted }

     % \abstract{}{}{}{}{} 
     % 5 {} token are mandatory
 
  \abstract
   {Extragalactic globular clusters  have been studied in elliptical galaxies 
     and in a few luminous spiral galaxies, but little is known about 
     globular clusters  in low-luminosity spirals.   }
   { Past observations with the ACS have shown that NGC~45  hosts a large population of globular
     clusters (19), as well as several young  star clusters. In this work we aim 
     to confirm the bona fide globular cluster status for 8 of 19 globular cluster candidates and to
     derive metallicities, ages, and velocities.}
   %methods
   {VLT/FORS2 multislit spectroscopy in combination with the Lick/IDS system was used to derive  
     velocities and to  constrain  metallicities and [$\alpha/$Fe] element ratio of the globular clusters.}
   {  We confirm the 8 globular clusters as bona fide globular clusters. Their velocities indicate halo or 
     bulge-like kinematics, with little or no overall rotation. From absorption
     indices such as H$\beta$, H$\gamma$, and H$\delta$  and the combined [MgFe]$'$ index, we found that 
     the globular clusters are metal-poor [Z/H]$\leq$-0.33 dex  and  [$\alpha/$Fe]$\leq$0.0 element ratio. 
     These results argue in favor of a population of globular clusters formed during the assembling of the galaxy.}
  {}

   \maketitle
%
%________________________________________________________________

%%%%%%%%%%%%%%%%%%%%%%%%%%%%%%%%%%%%%%%%%%%%%%%%%%%%%%%%%%%%%%%
\section{Introduction}
%%%%%%%%%%%%%%%%%%%%%%%%%%%%%%%%%%%%%%%%%%%%%%%%%%%%%%%%%%%%%%%

Globular clusters  are present in almost all kinds of galaxies.
Observations of  extragalactic globular cluster systems have shown that globular cluster systems can often be 
divided into (at least) two sub-populations, although the origin of these
remains unclear. In the Milky Way and M31, the metal-poor globular cluster sub-populations
display halo-like kinematics and spatial distributions \citep[e.g.][]{zinn85,ashmanzepf1998,Barmby2000,perrettM31}, while the
metal-rich globular clusters may be associated with the bulge and/or thick disk
\citep[e.g.][]{dante95,Barbuy98,cote1999,Bica2006}. 
The globular cluster sub-populations in elliptical galaxies show many similarities to
those in spirals, and some of the metal-rich clusters may have formed in
galaxy mergers \citep[e.g.][]{ashman1992}. Some of the metal-poor (``halo'')
clusters may have been accreted from dwarf galaxies 
\citep{dacosta} or in proto-galactic fragments from which 
the halo assembled \citep{searle}. However, there is evidence based on 
metallicity that globular clusters could not have been formed
from the destruction of dwarf galaxies \citep[see][and references there-in]{Koch2008}.
A major challenge is to establish how each one of these mechanisms may fit into the paradigm of hierarchical structure formation \citep[e.g.][]{santos}.

One important step towards understanding the roles of merging and accretion
processes is to extend our knowledge about globular clusters 
to many different galaxy types, such as dwarf galaxies and late-type spirals, in a range of environments.
Studies of globular clusters in spiral galaxies are more difficult than
in early-type galaxies because the globular cluster systems are generally poorer and
appear superposed on an irregular background. Consequently, most studies
of extragalactic globular clusters have focused on elliptical galaxies.

In spite of the similarities, there are also important differences between 
globular cluster systems of large ellipticals and those of spirals like the Milky Way.
Elliptical galaxies generally have many more globular clusters per unit 
host galaxy luminosity (i.e., higher globular cluster \emph{specific frequencies} \citet{harrisvandenbergh}
than 
spirals), and on average the globular cluster systems of ellipticals are more metal-rich 
\citep{markus1999}.
The best studied globular cluster systems in spiral galaxies are those associated with
the Milky Way and M31. Globular clusters in M31 appear similar to those in the Milky 
Way in terms of their  luminosity functions,  metallicities and, size distributions
\citep{cramptonM31,perrettM31,barmbyM31}.
M33 has a large number of star clusters 
\citep{christianM33_82,christianM33_88,chandar2001}
 but many of them have young ages and there may be
only a dozen or so truly ``halo'' globular clusters \citep{sarajedini2000}. 
The Magellanic Clouds are also well-known for their rich cluster systems,
but again only few of these are truly old globular clusters. The Large Magellanic Cloud
has about 13 old globular clusters \citep{johnson99LMC},
 which however show disk-like 
kinematics. The Small Magellanic Cloud has only one old globular cluster, NGC~121.

The globular cluster populations in low-luminosity spirals are almost unknown.
The question of how these clusters formed (and their host galaxy) remain
unanswered. Considering that these kind of galaxies remain almost unperturbed
during their live, it is probable that we are observing their first population
of globular clusters and therefore, we can approach to the conditions in which the host
galaxies were formed.

The nearby Sculptor group hosts several late-type galaxies whose globular
cluster systems are potentially within reach of spectroscopic observations
with 8 m telescopes in a few hours of integration time.
A previous study of the star cluster population in the Sculptor group was
done by \citet{knut}. They observed several globular clusters candidates, 
finding 19  globular clusters in four galaxies, most of them metal-poor with [$\alpha/$Fe] lower
than the Milky Way globular clusters.

In this paper we concentrate on the late-type, low-luminosity spiral
galaxy NGC 45 in Sculptor, in which we have previously identified a
surprisingly rich population of old globular cluster candidates in
HST/ACS imaging.
NGC~45 is classified as a  low-luminosity spiral galaxy with  
$B=11.37 \pm 0.11$ and $B-V=0.71$ \citep{paturel}. It is located at   
$\sim 5$ Mpc from us, $(m-M)_0=28.42 \pm 0.41$ \citep{bottinelli}, in the
periphery of the Sculptor group.
In \citet{Mora} we found 19 globular clusters
located in projection with the galaxy bulge, which appear to belong to the 
metal-poor population. Those 19 globular clusters yield a $S_N$ of $1.4-1.9$, which is high
for a late-type galaxy. 

In this paper we focus on 8 of those 19 globular clusters. We analyze them
through spectroscopy to confirm or reject their globular cluster status and
to constraint ages and metallicities.

%%%%%%%%%%%%%%%%%%%%%%%%%%%%%%%%%%%%%%%%%%%%%%%%%%%%%%%%%%%%%%%
\section{Candidates selection, observation and, reductions}
%%%%%%%%%%%%%%%%%%%%%%%%%%%%%%%%%%%%%%%%%%%%%%%%%%%%%%%%%%%%%%%

\subsection{Globular cluster selection}

  In \citet{Mora} we identified cluster candidates as extended
  objects, using a variety of size selection criteria based on the
  BAOLAB/ISHAPE \citep{larsen99} and SExtractor \citep{BERTIN}
  packages. We found  19  extended objects fulfilling the color criteria 
  $0.8<V-I<1.2$, with magnitudes  $V=19.5$ up to $V=22.5 $  that were interpreted 
  as globular clusters. The detected globular cluster candidates had a mean
  color of $V-I$ = 0.90, consistent with a metal-poor, old population.
  The mean half-light radius of the globular cluster candidates was
  $R_{\rm eff} = 2.9\pm0.7$ pc (error is the standard error of the mean), 
  similar to that of globular clusters in other galaxies.

\subsection{Observations and reductions}

\begin{figure}[h!]
  \centering
  \includegraphics[width=8cm]{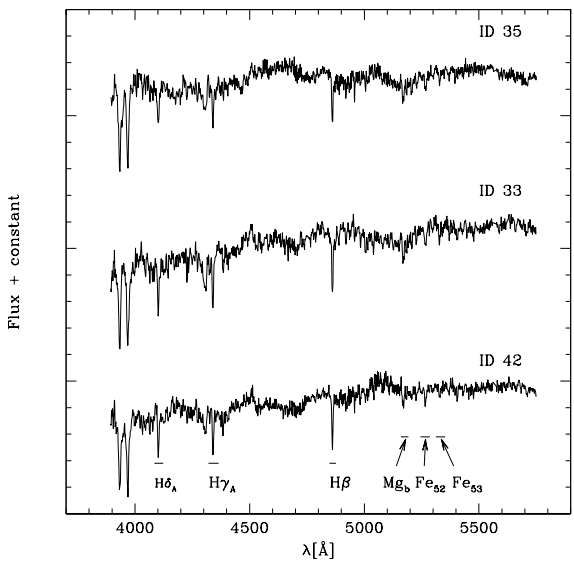}
  \caption
      {
	Samples of the spectra. The spectra have been shifted to the 0 radial
	velocity and an offset in flux has been added for best clarity of the sample. Some Lick-index passband are indicated at the bottom of the panel.
      }
          \label{template}
  \end{figure}

The spectra were acquired in service mode on the ESO period 77  using the ESO Focal Reducer/low
dispersion Spectrograph (FORS2) through the Multi-object spectroscopy mask exchange unit (MXU), 
which is mounted in Kueyen/UT2 VLT telescope at Cerro Paranal Chile. We used the GRIS\_600B$+$22 which has 
a wavelength range from 3330 \AA~up to 6210 \AA~with a dispersion at the central wavelength (4650 \AA)
of  0.75 \AA/pixel. Because of the globular cluster positions,  we were only able to place 8 globular clusters on the MXU.  Extra 
16 filler slits were placed on young star regions, the galaxy bulge, the sky and one star.
On each object a slit of 1.0$\arcsec$ width was placed and MXU observations were done in 7 Observing Blocks (OBs). A sample of these spectra are shown in Fig. \ref{template}.
For the calibration, we acquired 6 Lick/IDS standard stars from \citet{Lick_standards} in  long slit spectra mode using the 
same configuration of the globular cluster observations (for a log of the observations see Table \ref{LOG_OBS}).
The six standard stars were selected according to their spectral type (i.e. K3III, G0V, G5IV, F5VI, K3IIIV and, A3V), covering the range of spectral types expected in a globular cluster. Each standard was observed in 3 
exposure sets with a small offset in the direction of the slit in order to avoid bad pixels in the final 
combined spectra.

\begin{table}[h!]
\caption[]{Log of the observations. 
}
\label{LOG_OBS}
$$
\begin{array}{p{0.18\linewidth}rrcc}
\hline
\hline
Object &\mathrm{Mode}  & \mathrm{N~of~exposures} \times \mathrm{time}  &   \mathrm{Obs~type} \\
\hline
NGC~45   & \mathrm{MXU} & 17\times(3\times1200s)  & \mathrm{Science} \\
NGC~45   & \mathrm{MXU} & 2\times1095s            & \mathrm{Science}\\
HD180928 & \mathrm{LSS} & 3\times 1.00s           & \mathrm{Lick~std} \\
HD195633 & \mathrm{LSS} & 3\times 1.00s           & \mathrm{Lick~std} \\
HD165195 & \mathrm{LSS} & 2\times(3\times 3.00s)  & \mathrm{Lick~std} \\ 
HD003567 & \mathrm{LSS} & 3\times 5.00s           & \mathrm{Lick~std} \\
HD221148 & \mathrm{LSS} & 3\times 1.00s           & \mathrm{Lick~std} \\
HD006695 & \mathrm{LSS} & 2\times (3\times 0.63s)   & \mathrm{Lick~std} \\ 
\noalign{\smallskip}
     \hline 
   \end{array}
   $$
The second column indicates the observed mode: 
Multi-object spectroscopy mask exchange unit (MXU) and Long slit spectroscopy (LSS).
\end{table}

The spectra (science and standard stars) were reduced (i.e. bias subtracted, flat field corrected, 
optical distortion corrected and, wavelength calibrated) using the ESO Recipe Execution Tool v3.6 
(ESO-REX)\footnote{http://www.eso.org/sci/data-processing/software/cpl/esorex.html}. Typical 
rms from the distortion  corrections were on the order of 0.4 pixels and 
the wavelength  calibration accuracy of the model applied during the wavelength calibration was on the order of 0.08\AA.

In the following section we explain the radial velocity measurements. 
We only focus on the globular clusters because filler spectra were too faint 
to have reasonable radial velocity measurements.

%%%%%%%%%%%%%%%%%%%%%%%%%%%%%%%%%%%%%%%%%%%%%%%%%%%%%%%%%%%%%%%
\section{Radial velocities}
%%%%%%%%%%%%%%%%%%%%%%%%%%%%%%%%%%%%%%%%%%%%%%%%%%%%%%%%%%%%%%%

Radial velocities were first derived for the standard stars. Since each standard 
star was acquired in a set of 3 consecutive exposures, we derived the radial velocity
on each single exposure. This was accomplished by cross correlating a zero velocity  
elliptical galaxy template from \citet{quintana} with each  spectrum using the \emph{FXCOR} 
IRAF\footnote{IRAF is distributed by the National Optical Astronomical Observatory, which is 
operated by the Association of Universities for Research in Astronomy,  Inc, under  cooperative 
agreement with the National Science Foundation.}  task. Each single spectrum  was shifted 
to zero radial velocity and combined  into a high  signal to noise standard star spectrum.

Globular cluster spectra taken in the same OB were combined, yielding 
7 spectra for each globular cluster. Each one of them, in combination with 
the high signal to noise standard star spectra, were used to derive the globular cluster
velocities through cross correlation using the \emph{FXCOR} IRAF task. On each cross 
correlation we select a region of  200 \AA ~width  centered in the  Ca II H+K, H$\beta$ and, 
H$\gamma$  features. An  extra  region from 5000 \AA~ up to 5500 \AA~ was also 
considered for cross correlation. We  note that the cross correlations between  
standard-star types A and K; and the globular cluster spectra 
were particularly difficult, most likely because such stars provide a poorer
match to the overall spectrum of a globular cluster.  This effect, combined with the low
 signal-to-noise of the spectra (especially for the globular clusters ID 45 and ID 47)  
caused the velocity measurements to be more uncertain when based on these stars.

\begin{figure}[h!]
  \centering
  \includegraphics[width=8cm]{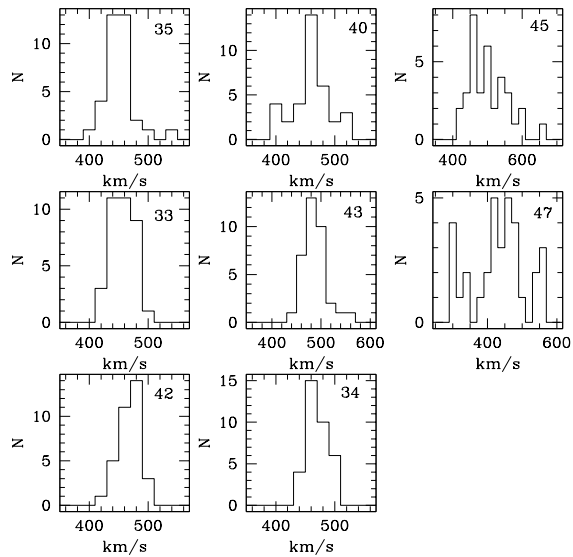}
  \caption
      {
	Velocity distributions of the globular clusters. The bin size is 20 km/sec. The number on each pannel corresponds to the globular cluster ID. 
      }
 \label{velfig}

  \end{figure}

In Fig. \ref{velfig} we show histograms of all the individual  velocity measurements for each cluster.
 We adopted the average of 
the distribution as the final velocity value of each  globular cluster. The error
was obtained from the standard deviation divided by the square root of the number of measurements. 
Values are listed in Table \ref{tabla_velocidades}.

\begin{figure}[h!]
  \centering
  \includegraphics[width=9cm]{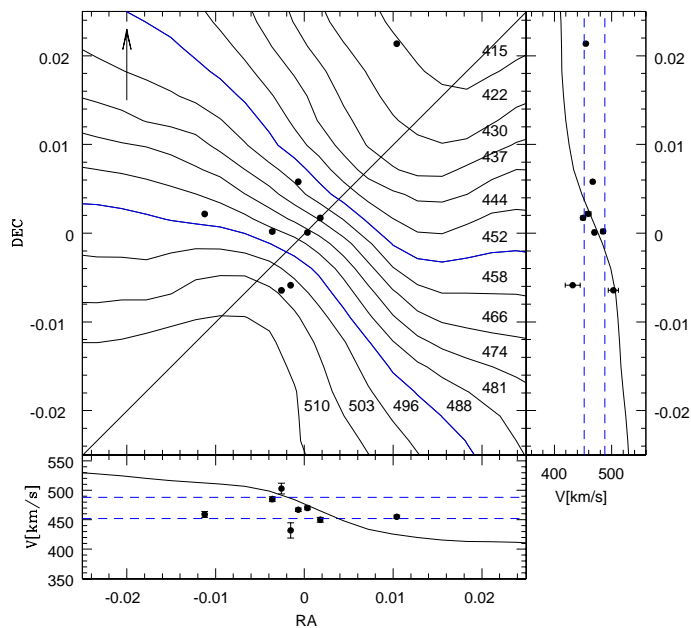}
  \caption
      {
	Main panel: Position of the bona fide globular clusters with 
	overplotted  isovelocity contours from \citet{chemin}.
	Arrow indicates the north and East is on the left.
	The sub panels on the bottom as well as on the right 
	show the globular cluster velocity projections of the main 
	panel as function of RA and DEC respectively. Blue lines indicate
	isovelocity lines for 480 km/s and 452 km/s. The straight line crossing 
	the main panel has the purpose of illustrating the velocity changes 
	when it is projected as function of RA and DEC on the sub panels. 
	Numbers next to the isovelocity contours indicate the velocity in km/s.
      }
      
          \label{VELOCIDADES}
  \end{figure}

\begin{table}[h!]
\caption[]{ Derived velocities of the globular clusters. 
}
\label{tabla_velocidades}
$$
\begin{array}{cccc}           
   \hline                                     
   \hline              
   \noalign{\smallskip}
%   (1)&(2)&(3)\\
   \mathrm{ID} & \mathrm{Vel(km/s)} &\sigma & \mathrm{N} \\
   \hline              

33 & 455 \pm3  & 19 & 37\\
34 & 470 \pm3  & 17 & 35\\
35 & 450 \pm4  & 25 & 35\\
40 & 459 \pm5  & 32 & 35\\
42 & 467 \pm3  & 17 & 34\\
43 & 485 \pm4  & 23 & 35\\
45 & 503 \pm9  & 55 & 33\\
47 & 432 \pm13 & 77 & 33\\

\noalign{\smallskip}
     \hline
   \end{array}
   $$
$\sigma$ is the standard deviation and N corresponds to the number of measurements.
\end{table}
In Fig. \ref{VELOCIDADES} we show the position of the globular clusters with overplotted
isovelocity contours from \citet{chemin}. The sub-panel on the bottom and the sub-panel 
on the right shows the projected  velocity as function of RA and DEC. The panels show 
that our velocity measurements are consistent with no overall rotation of the globular
cluster system. Globular clusters located near the center of the galaxy show velocities 
consistent with the observed H~I gas velocities from \citet{chemin}. The greatest difference
between the globular cluster  velocities was  ${\Delta}\mathrm{V} =71 \pm 16 $ km/s. It
corresponds to the difference of velocity between the globular cluster ID=45 and 47, 
which also show the largest errors. Therefore, globular cluster velocities are  mainly 
concentrated between V=430 -- 480 km/s, as is seen in Fig \ref{VELOCIDADES}.

The velocities of the GCs clearly do not match the isovelocity
contours of \citet{chemin}, and we thus exclude that the GCs
are associated with the disk component of NGC~45.  The velocity dispersion
is $\sigma = 20\pm4$ km/s, which is significantly larger than the
measurement errors (Table \ref{tabla_velocidades}) and smaller than other similar galaxies like M~33 (\citet{chandar2002} found values of $\sigma=54\pm8$ for 
disk/bulge globular clusters and $\sigma=83 \pm13$ km/s  for halo globular clusters).
Therefore, we conclude that the globular
cluster velocities in NGC~45 are indicative of halo- or bulge like
kinematics, with little or no overall rotation.

%%%%%%%%%%%%%%%%%%%%%%%%%%%%%%%%%%%%%%%%%%%%%%%%%%%%%%%%%%%%%%%
\section{Lick index calibrations}
%%%%%%%%%%%%%%%%%%%%%%%%%%%%%%%%%%%%%%%%%%%%%%%%%%%%%%%%%%%%%%%

In the following we estimate abundances for our globular cluster sample using the
Lick/IDS system of absorption line indices. Passband definitions were taken 
from G.~Worthey's web page\footnote{http://astro.wsu.edu/worthey/html/index.table.html} which corresponds to 
\citet{Trager98}  for the new wavelength definitions, \citet{Lick_standards} for the old definitions 
including  H$_\delta$ and H$_\gamma$ definitions from \citet{Lick_resolution}.  

The Lick/IDS system is designed to measure absorption features such as CN, H$\beta$, Fe, Mg, G (molecular
bands) and blend of absorption lines present in old populations. These features were used to construct a library
from several stars observed at the Lick observatory.  Six of these standard stars were taken
during the observations that are presented in this work.    
Due to  our telescope configuration, the  standard stars spectra, as well as the globular cluster spectra, 
have  higher spectral resolution  than the original Lick/IDS system. Thus,  we must carefully degrade our 
spectra in order to match the Lick/IDS spectral resolution. One way to quantify the difference 
between our instrumental system and the original Lick/IDS spectra is to measure the FWHM of narrow spectral 
features. We measured the FWHM of the sky lines in our sky spectra and we found a typical
FWHM=4.9 \AA.~  This value was used as input for the code used for the index derivations.

The indices were measured using the GONZO code from \citet{Thomas_GONZO}. Briefly, GONZO degrades the spectra 
to the Lick/IDS index system with a wavelength-dependent Gaussian kernel in order to match
the resolution from \citet{Lick_resolution}. GONZO also  derives the uncertainties  of the indices 
by considering  the Poisson statistics from the error spectra. These statistics are used in addition to the 
random noise when creating artificial science spectra and measuring the indices on these. 
For further details on GONZO, see  \citet{Thomas_GONZO}.

\begin{table}[h!]
\caption[]{Calibration summary of the Lick Indices. }
%TiO$_1$ and TiO$_2$ were not 
%measured due to the  fact that our standard-star spectra  do not cover the TiOs wavelength. 
\label{calibrationW}
$$
\begin{array}{p{0.18\linewidth}rrc}           
   \hline                                     
   \hline              
   \noalign{\smallskip}
%   (1)&(2)&(3)&(4)\\
   {Index} & \mathrm{ZP} & \sigma & \mathrm{Units} \\
\hline
CN$_1$          &   0.0301     &    0.0123      &    mag\\
CN$_2$          &   0.0364     &    0.0087      &    mag\\
Ca4227          &   0.3878     &    0.0830      &    \AA\\
G4300           &   0.8417     &    0.2467      &    \AA\\
Fe4383          &   0.6260     &    0.2155      &    \AA\\
Ca4455          &   0.1553     &    0.1183      &    \AA\\
Fe4531          &   0.3835     &    0.2359      &    \AA\\
Fe4668          &  -0.0715     &    0.2840      &    \AA\\
H$\beta$        &   0.0515     &    0.1234      &    \AA\\
Fe5015          &   0.0583     &    0.2394      &    \AA\\
Mg$_1$          &  -0.0116     &    0.0034      &    mag\\
Mg$_2$          &  -0.0149     &    0.0062      &    mag\\
Mg$_b$          &   0.2932     &    0.0501      &    \AA\\
Fe5270          &   0.4598     &    0.0563      &    \AA\\
Fe5335          &   0.2400     &    0.1041      &    \AA\\
Fe5406          &   0.0729     &    0.0581      &    \AA\\
Fe5709          &   0.1342     &    0.0667      &    \AA\\
TiO$_1$         &    -	       &  	-	&    mag\\	  
TiO$_2$         &    -	       &  	-	&    mag\\	  
H$\delta_A$     &   0.1376     &    0.3270	&    \AA\\
H$\gamma_A$     &  -0.4629     &    0.3295 	&    \AA\\
H$\delta_F$     &   0.6445     &    0.1504      &    \AA\\
H$\gamma_F$     &  -0.1960     &    0.1333      &    \AA\\

\noalign{\smallskip}
     \hline
   \end{array}
   $$
\end{table}

In the comparisons between our standard stars and the Lick/IDS system, we assumed 
that the transformation between our measured Lick/IDS
indices and the standard values was linear with a slope of unity,
so that only an offset is needed to match the standard Lick/IDS
system. Also  we  avoided possible changes of the slopes  due to outlier measurements 
(which were  impossible to identify because of  the small number of the observed standard stars). 

\begin{figure*}[]
  \centering
  
  \includegraphics[width=15cm]{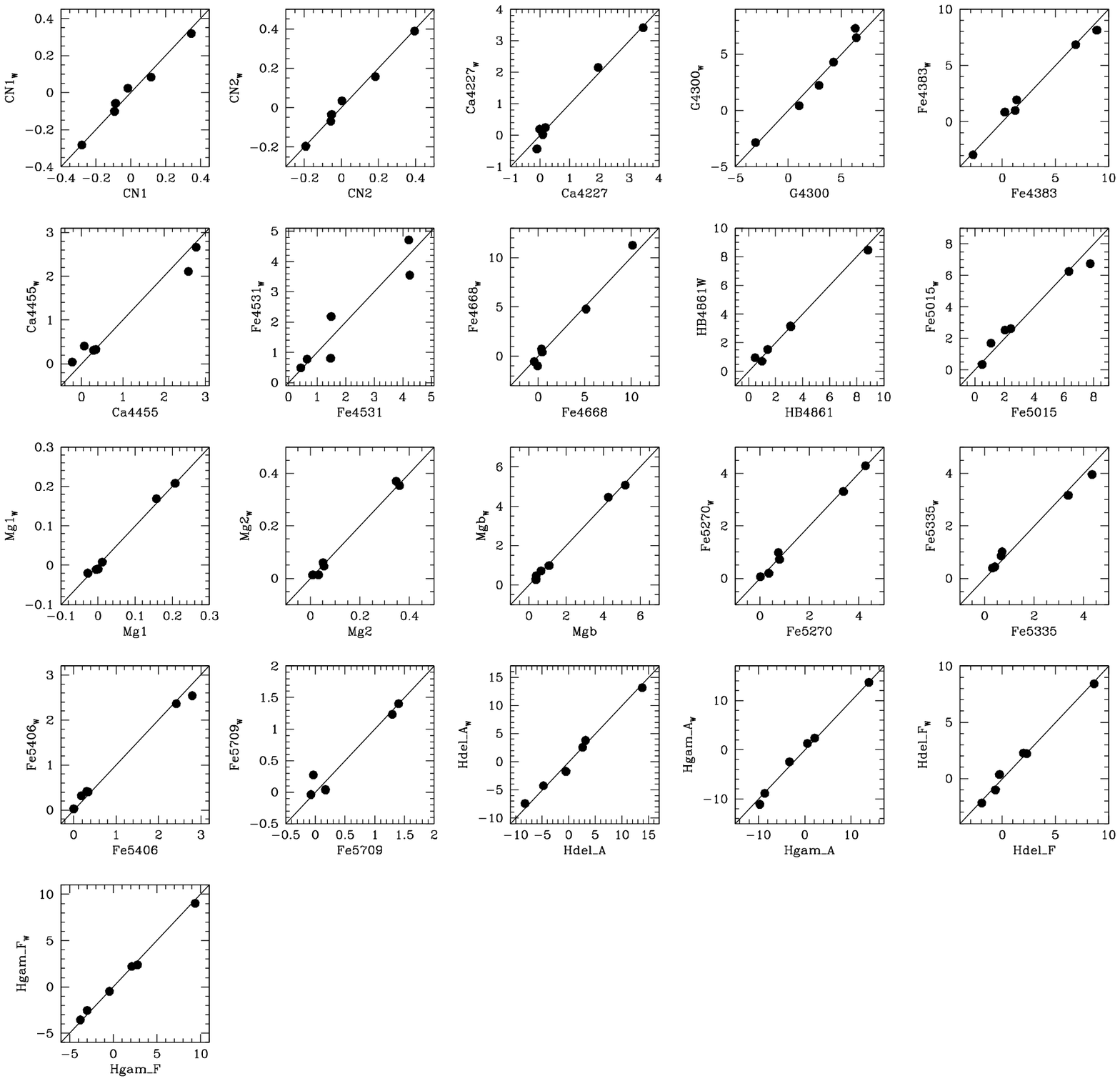}
  \caption
      {
	Comparison of Worthey Lick/IDS indices with the one measured by us. 
	Lines represent the one-to-one comparison. 
      }
          \label{lickcomparison}
 \end{figure*}

The zero-point offsets are given in Table \ref{calibrationW} and a comparison
between our measurements and the Lick/IDS system is shown in Fig. \ref{lickcomparison}. 
In Table \ref{calibrationW} ZP corresponds  to the zero point needed in order to match 
the Lick/IDS system and $\sigma$  corresponds to the error which was calculated considering
the standard deviation divided by the square root of the number of measurements.
The reader may note that we do not list values for the TiO indices because
the TiO indices were outside the wavelength coverage.

%%%%%%%%%%%%%%%%%%%%%%%%%%%%%%%%%%%%%%%%%%%%%%%%%%%%%%%%%%%%%%%
\section{Results} 
%%%%%%%%%%%%%%%%%%%%%%%%%%%%%%%%%%%%%%%%%%%%%%%%%%%%%%%%%%%%%%%

%%%%%%%%%%%%%%%%%%%%%%%%%%%%%%%%%%%%%%%%%%%%%%%%%%%%%%%%%%%%%%%
	\subsection{Age diagnostic plots}
%%%%%%%%%%%%%%%%%%%%%%%%%%%%%%%%%%%%%%%%%%%%%%%%%%%%%%%%%%%%%%%

In this subsection we discuss the results of the indices measurements and their comparisons with 
$\alpha/$Fe-enhanced  models from  \citet{Thomas03} and \citet{Thomas04}. We adopted the index 
$[\mathrm{MgFe}]'=\sqrt{\mathrm{Mgb}\times(0.72\times \mathrm{Fe5270} + 0.28 \times \mathrm{Fe5335)}}$
which is [$\alpha/$Fe] independent  defined by \citet{Thomas03}. We also  adopted from \citet{gonzalez}
the $<\mathrm{Fe}>=(\mathrm{Fe5270}+\mathrm{Fe5335})/2$   index definition.
\begin{figure*}
  \centering
  \includegraphics[width=5.6cm]{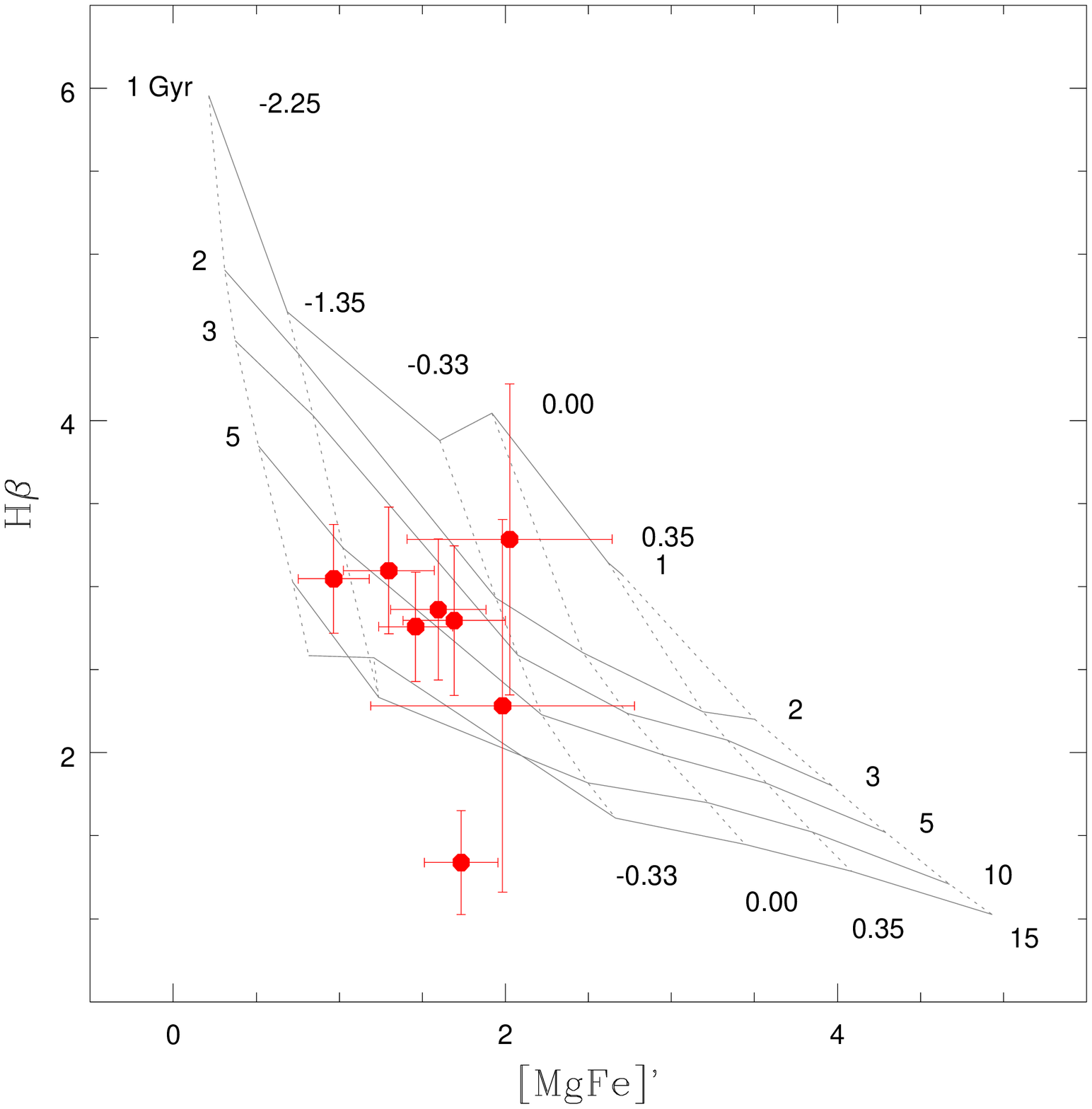}
  \includegraphics[width=5.6cm]{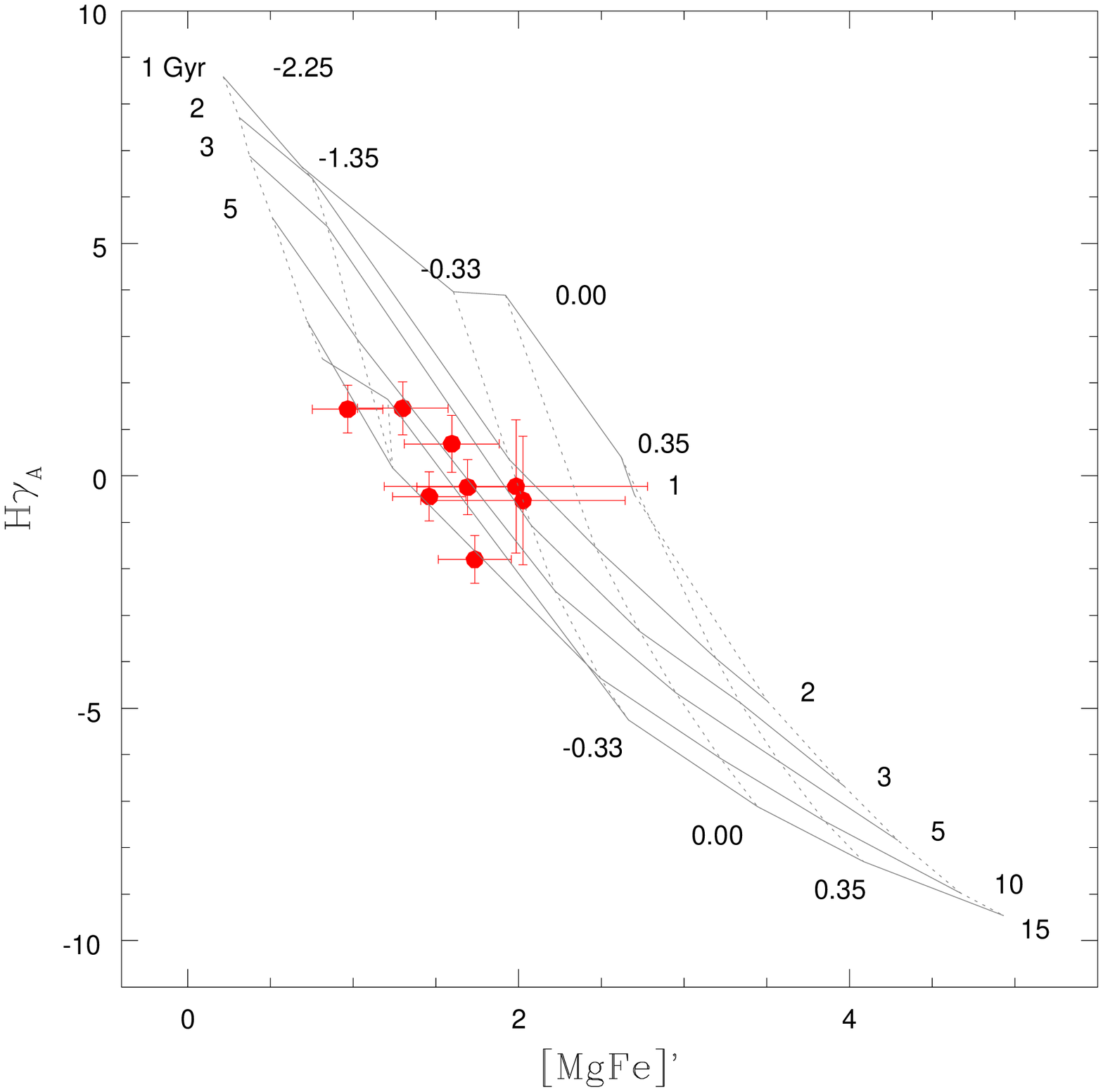}
  \includegraphics[width=5.6cm]{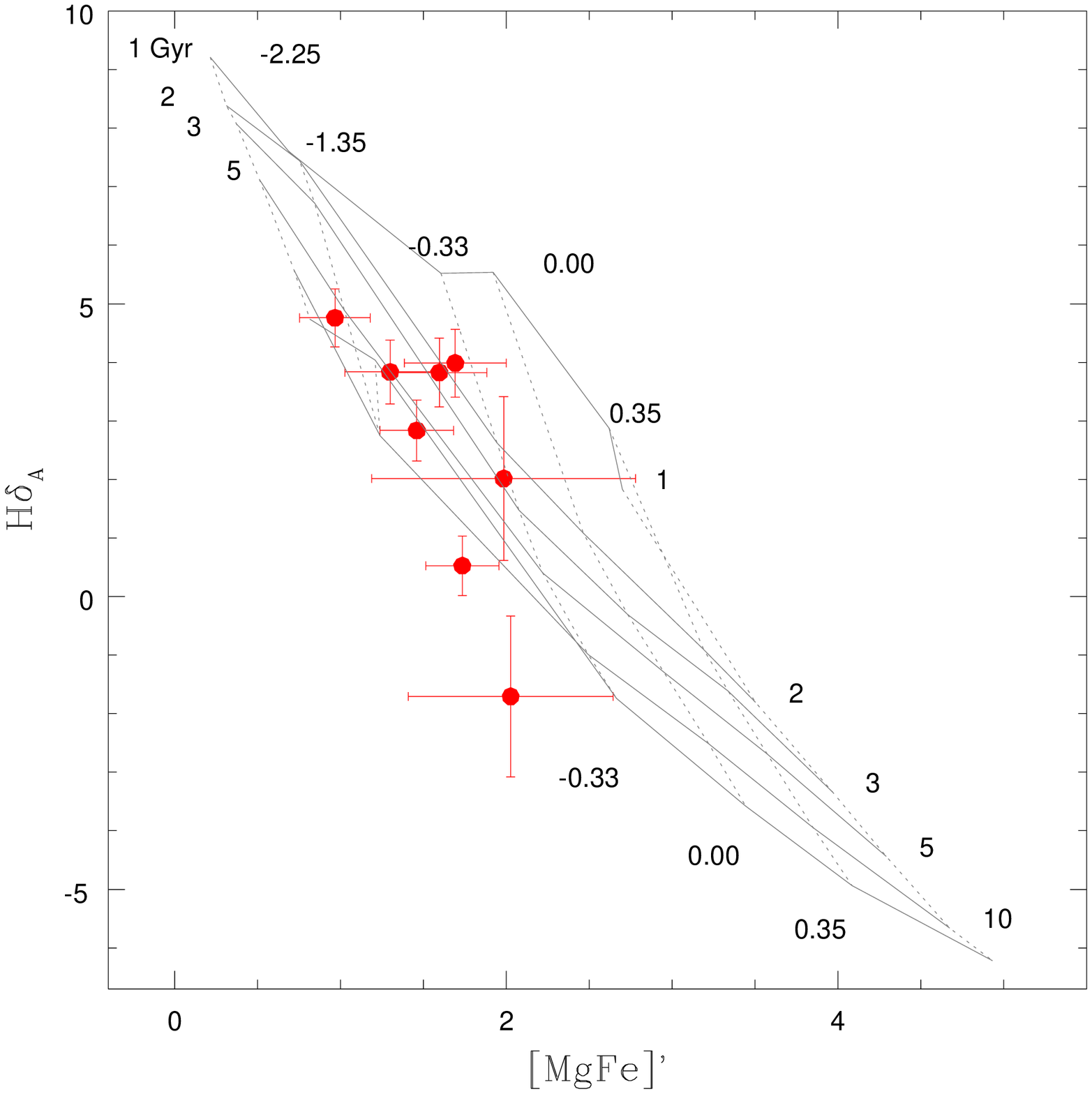}
  \includegraphics[width=5.6cm]{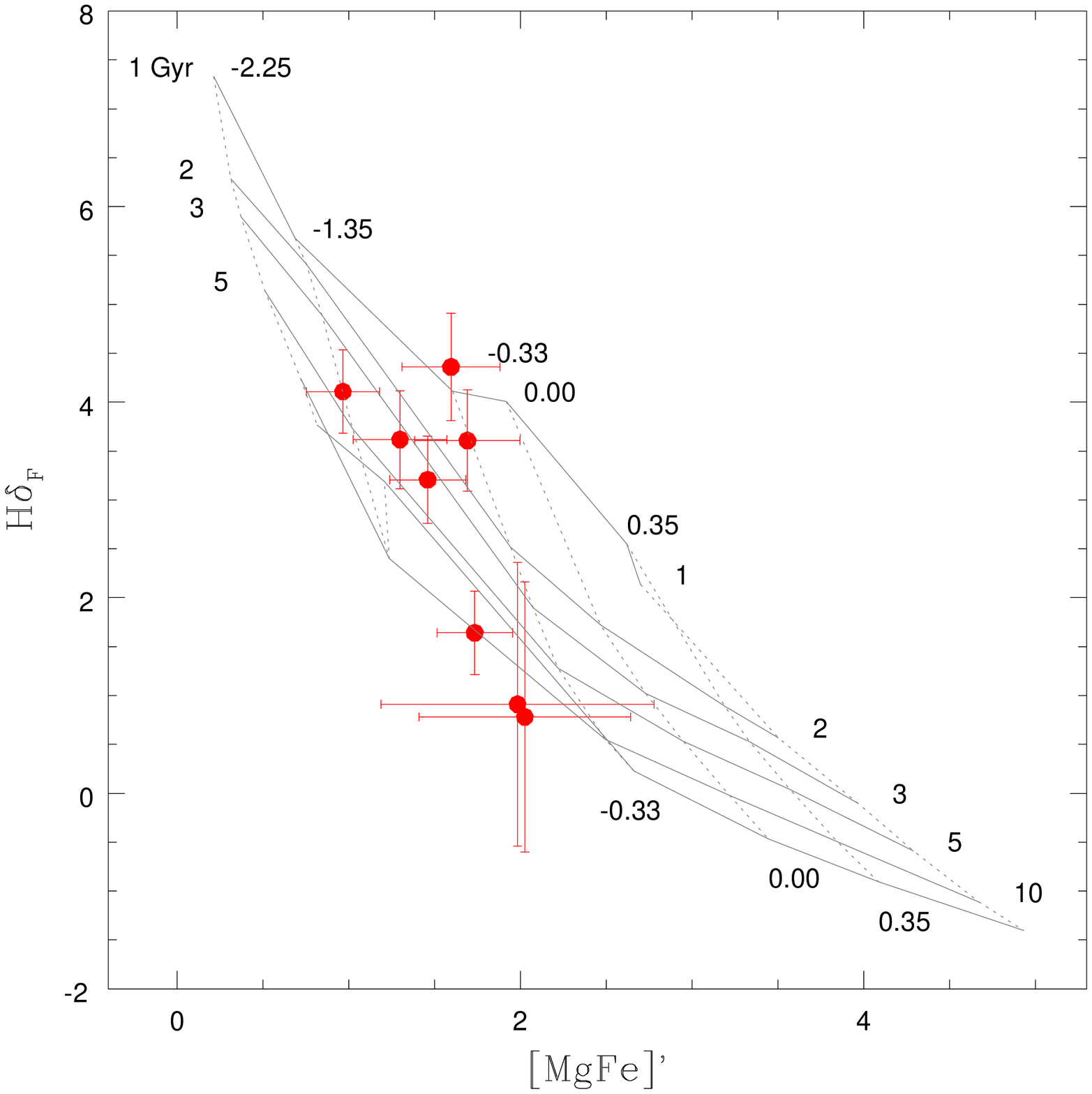}
  \includegraphics[width=5.6cm]{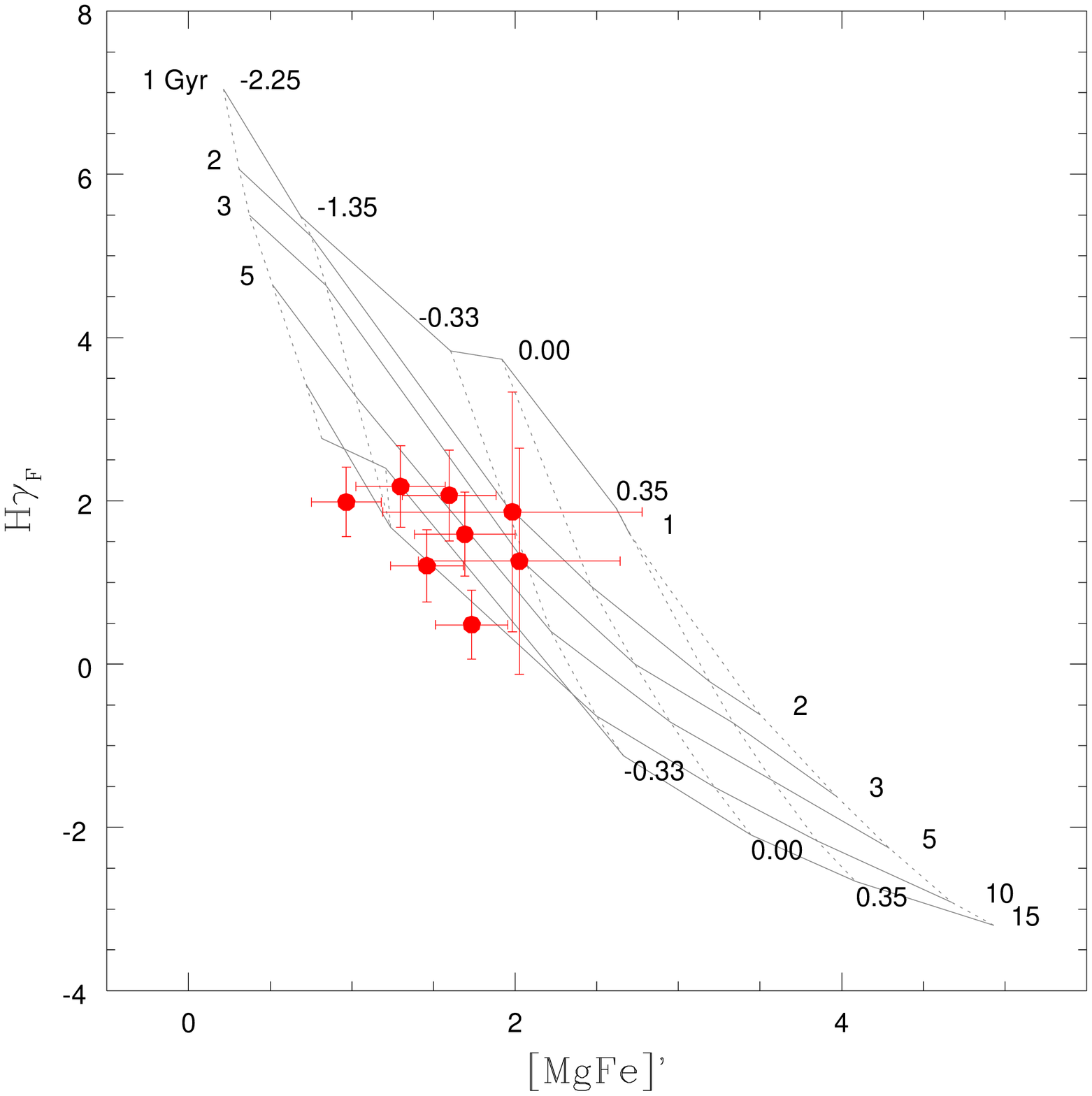}
  \caption
      {
	Age diagnostic plots. The over-plotted grid corresponds to SSP model from \citet{Thomas04}
	for [$\alpha/$Fe]$=0.0$. Doted lines correspond to  metallicities 
	[Z/H]= $-2.25$, $-1.35$, $-0.33$, 0.0, 0.35 and, 0.67 dex.
	Solid lines correspond to  ages of  1, 2, 3, 5, 10 and 15 Gyr.      
      }
          \label{Grids}
  \end{figure*}
Figure \ref{Grids} show the age metallicity diagnostic plots  for the 
Balmer line indices H$\beta$, H$\gamma_A$, H$\gamma_F$, H$\delta_A$ and, H$\delta_F$ against [MgFe]$'$. 
All our globular clusters show [MgFe]$'$ values less than or equal to 2 \AA, which makes them metal poor.

\begin{figure*}
  \centering
  \includegraphics[width=5.8cm]{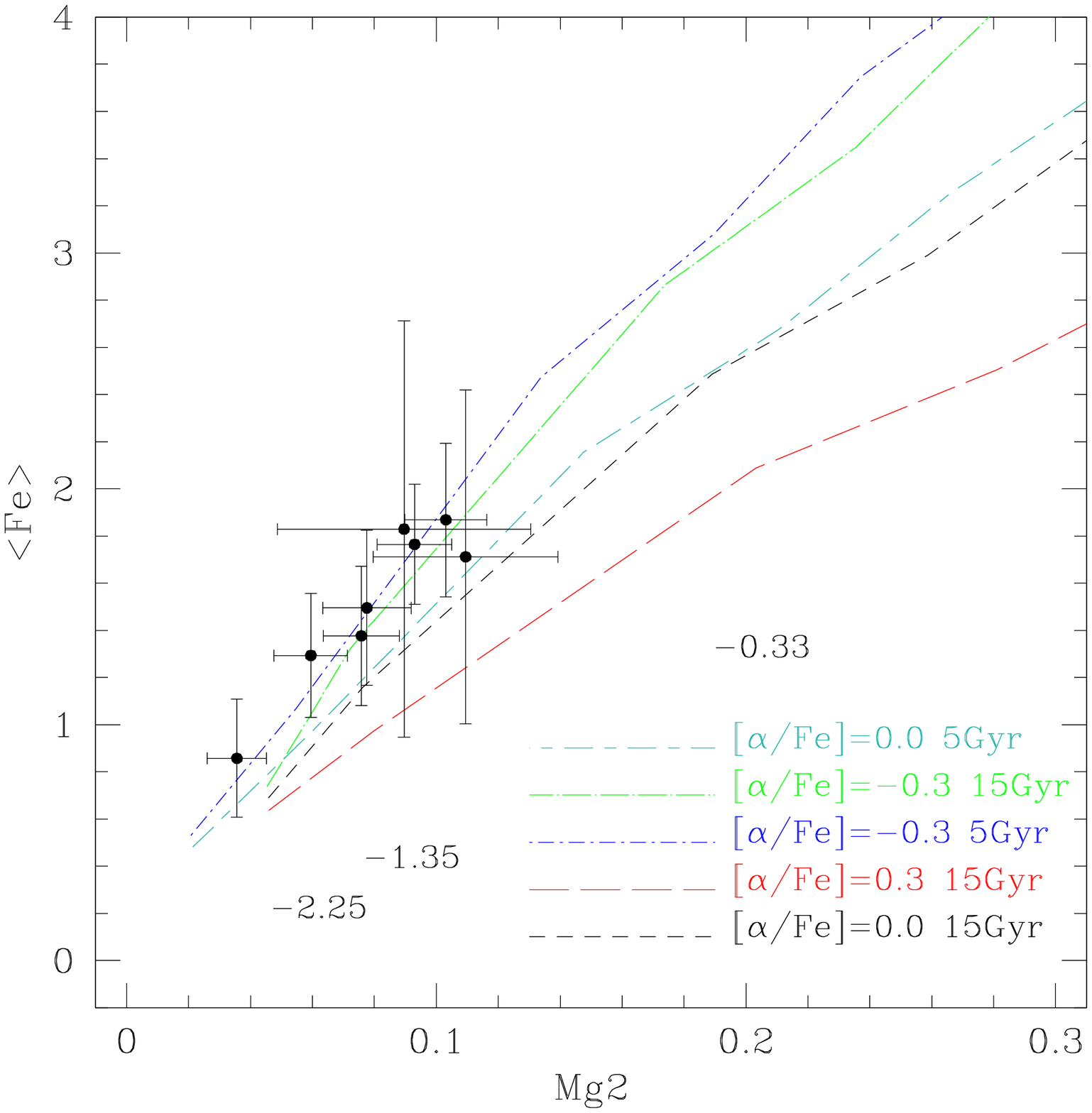}
  \includegraphics[width=5.8cm]{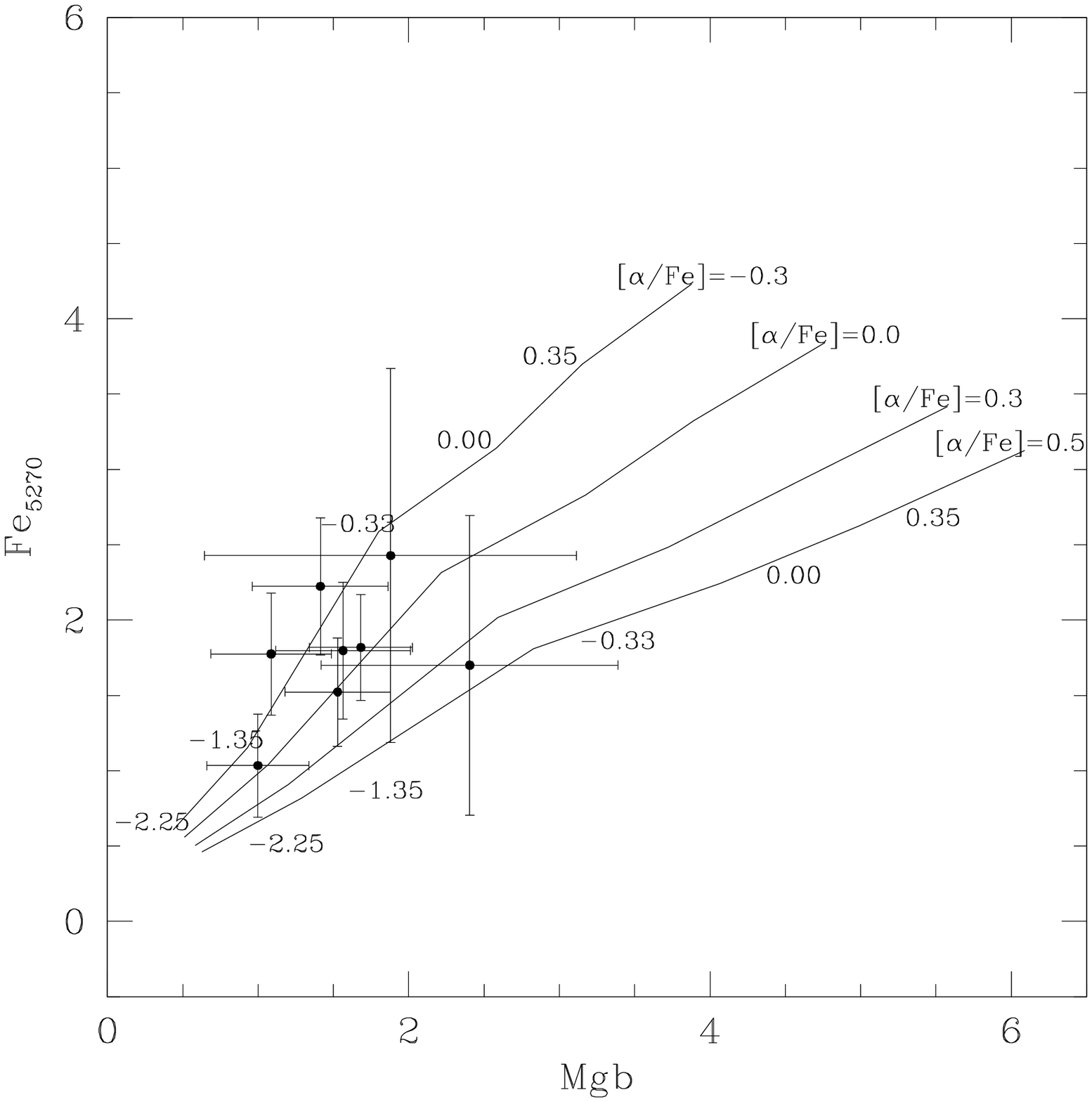}
  \includegraphics[width=5.8cm]{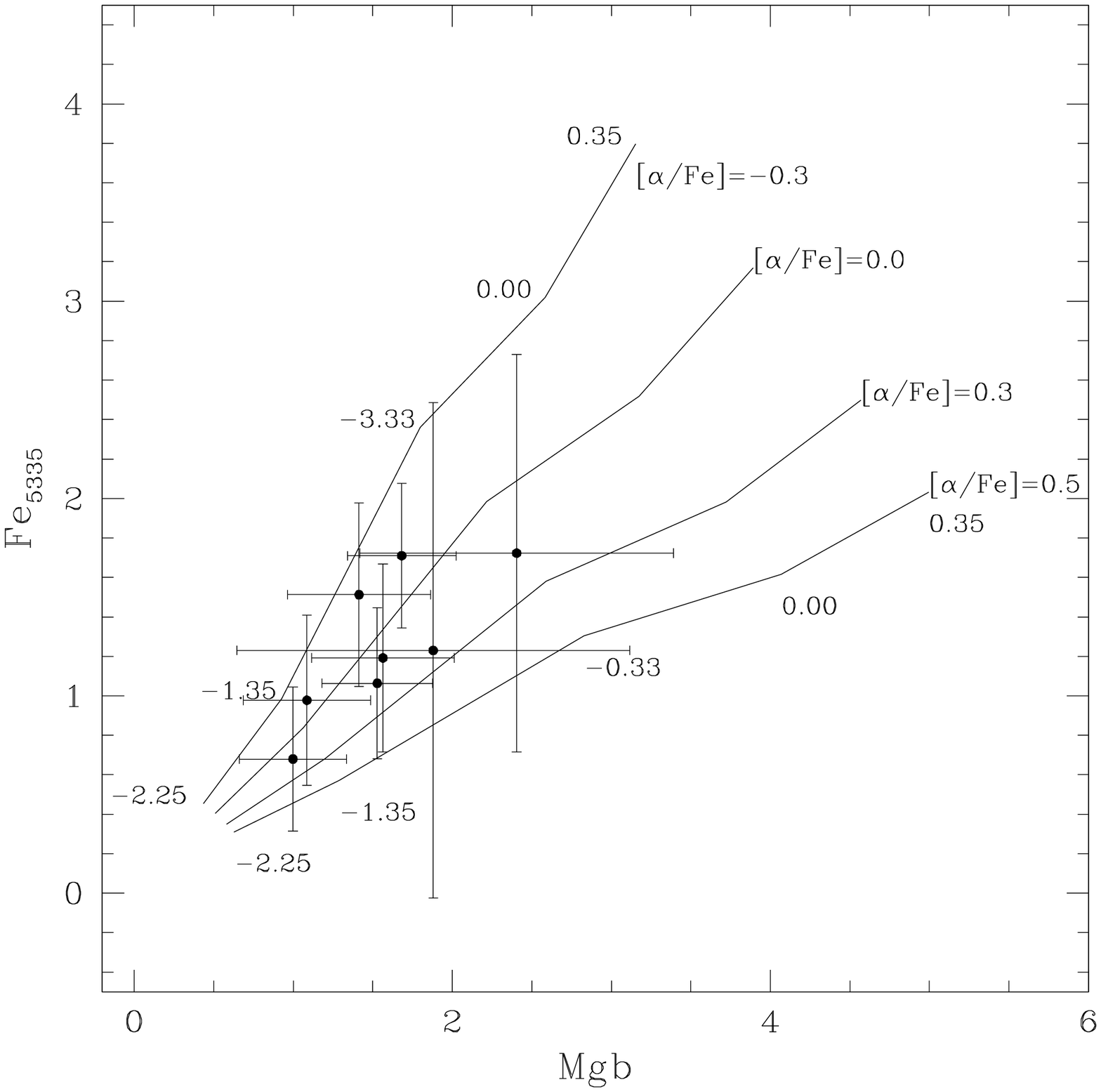}

  \caption
      {
	Left panel: Iron $<$Fe$>$ as function of Mg2.
	Central panel:  Fe$_{5278}$ index versus Mgb.
      	Right panel:  Fe$_{5335}$ index versus Mgb.
      }

      \label{Grids2}
\end{figure*}

Ages are poorly constrained and depend on the Balmer line used
for the comparison with models, but the metallicities are
consistently sub-solar with [Z/H]$<$-0.33. 
In the left panel of  Fig. \ref{Grids2} we show the average iron versus Mg2 over-plotted with models from \citet{Thomas04} 
for [$\alpha/$Fe]=$-0.3$, 0.0 and, 0.3; and ages of 5 and 15 Gyr. 
The panel shows globular clusters located in the sub-solar metallicity 
region with values lower than [Z/H]=$-0.33$ corroborating the same result of Fig. \ref{Grids} 
and, despite the great uncertainties, the globular clusters seem to be better described by 
models with sub-solar [$\alpha/$Fe]=0.0, $-0.3$ element ratios. 
This conclusion becomes less strong when the two Fe index are plotted separately 
versus Mgb  as it is seen in the central and right panels of Fig. \ref{Grids2}.  For the Fe$_{5270}$ index,  globular clusters 
show [$\alpha$/Fe]$ \leq 0.0$ while for the Fe$_{5335}$ index,  globular clusters  are 
scattered between $-0.3 <$[$\alpha$/Fe]$<0.3$. The derived values of [$\alpha/$Fe] element 
ratios and metallicities  seems to be consistent with values found by \citet{knut} in
4 Sculptor galaxies.
They concluded that all globular clusters in the Sculptor group have [Fe/H]$\leq$ $-1.0$ 
(we found values for NGC~45 [Fe/H]$\leq$ -0.3) and values of 
$-0.3 \pm  0.15  <$[$\alpha/$Fe] $ < $ 0.0 $\pm$ 0.15 for the measured globular clusters.

%%%%%%%%%%%%%%%%%%%%%%%%%%%%%%%%%%%%%%%%%%%%%%%%%%%%%%%%%%%%%%%
\subsection{Comparison with photometric ages}
%%%%%%%%%%%%%%%%%%%%%%%%%%%%%%%%%%%%%%%%%%%%%%%%%%%%%%%%%%%%%%%

In spite of the uncertainties of the star cluster metallicities and ages, it is worth to comment
how ages and metallicities derived here compare with the previous ages and 
metallicities assumed in \citet{Mora}. As a reminder, in \citet{Mora}
we used GALEV models \citep{anders} considering 4 metallicites: Z=0.004, 0.008, 0.02, and 0.05
which are equivalent to [Fe/H]$=-0.7$, 0.4, 0, and $+0.4$. 
In the present work, we have 3 globular clusters (ID=33, 34 and, 35) in common with our previous work.

In Table \ref{Metal}  we show the  [Z/H] values calculated for each index and its mean value 
calculated from the age diagnostic plots. 
We do not extrapolate ages nor metallicities of globular clusters lying  outside the model grids. 
Errors were calculated considering the highest and lowest values for each index within the error bars. Also,  
if the error bar (or a part of it) lie outside the model grids, we do not extrapolate its value and we
considered the farest point of the grid as the maximum (or minimum) error value. Therefore, errors lie within
the grid values. Globular clusters where this happened are shown in boldface on the Table \ref{Metal}. We found that the derived metallicities agree on each  index and we concluded that the
 most adequate metallicity for age and mass derivations with photometry is Z=0.004 of the GALEV models.

\begin{table}[h!]
\caption[Metallicities from age diagnostic plots.]{Metallicities from age diagnostic plots. 
}

\label{Metal}
$$
\tiny
\begin{array}{ccccccc}           
   \hline                                     
   \hline              
   \noalign{\smallskip}
 (1)&(2)&(3)&(4)&(5)&(6) &(7)\\
 \mathrm{ID}  &  \mathrm{Z/H} (H_{\beta})  &  \mathrm{{Z/H~}}(H\gamma_{\mathrm{A}}) & \mathrm{{Z/H~}}(H\delta_{\mathrm{F}}) & \mathrm{{Z/H~}}(H\delta_{\mathrm{A}}) & \mathrm{Z/H~}(H\gamma_{\mathrm{F}}) & <\mathrm{[Z/H]}>\\

\hline
33 & -1.1 ^{+0.40}_{-0.55}      & \bf{-1.15^{+0.45}_{-0.60}}   & -1.0^{+0.65}_{-0.75}        & \bf{-1.15^{+0.60}_{-0.55}}   & \bf{-1.2 ^{+0.40}_{-0.50}}   &-1.12^{+0.5}_{-0.59}\\ 
34 &          -                          & \bf{-0.9 ^{+0.2 }_{-0.10}}    &                   -                 &                  -                     &        -                        &-0.9^{+0.2}_{-0.1}   \\  
35 & -0.85^{+0.45}_{-0.40}      & -0.8 ^{+0.3 }_{-0.50}          &                  -                  & -0.65^{+0.30}_{-0.55}         & -0.85^{+0.40}_{-0.40}  &-0.79^{+0.36}_{-0.46}\\ 
\hline		     	     				    		    	  	   	    			 		      
40 & -1.6 ^{+0.45}_{-0.65}       &                    -                  & \bf{-1.4^{+0.45}_{-0.85}}  & \bf{-1.6 ^{+0.55}_{-0.65}}     &        -                        &-1.53^{+0.48}_{-0.72}\\ 
42 & -1.0 ^{+0.35}_{-0.35}       & \bf{-1.15^{+0.35}_{-0.15}}  & \bf{-0.9^{+0.55}_{-0.45}}  & \bf{-1.05^{+0.40}_{-0.30}}    &        -                        &-1.03^{+0.41}_{-0.31}\\ 
43 & -0.75^{+0.40}_{-0.45}       & \bf{-0.8 ^{+0.40}_{-0.40}}  & -0.4^{+0.30}_{-0.60}        & -0.45^{+0.30}_{-0.60}          & \bf{-0.8 ^{+0.40}_{-0.40}}   &-0.64^{+0.36}_{-0.49}\\ 
45 & \bf{-0.15^{+0.65}_{-1.05}} & \bf{-0.35^{+0.95}_{-0.85}}  &                   -                 &                 -                       & \bf{-0.4 ^{+0.70}_{-0.80}}    &-0.30^{+0.76}_{-0.9}\\ 
47 & -0.6 ^{+0.65}_{-0.80}        & \bf{-0.35^{+0.60}_{-1.05}}  &                   -                 & \bf{-0.35^{+0.55}_{-1.00}}    & \bf{-0.3 ^{+0.55}_{-1.10}}   &-0.40^{+0.59}_{-0.99}\\ 

\noalign{\smallskip}
     \hline
   \end{array}
   $$
Column (1):  Globular  cluster ID. Columns (2)  to   (6):  Z/H. The index used for its derivation is shown between brackets. 
Column (7): average Z/H

\end{table}

\begin{table}[h!]
\caption[Ages from diagnostic plots.]{Derived ages in Gyr from age diagnostic plots. 
}
\label{Edades}
$$
\tiny
\begin{array}{ccccccccc}           
   \hline                                     
   \hline              
   \noalign{\smallskip}
   (1)&(2)&(3)&(4)&(5)&(6) & (7) & (8) &(9)\\
 \mathrm{ID}  &  \mathrm{Age} (\mathrm{H}_{\beta})  &  \mathrm{Age} (\mathrm{H}\gamma_{A}) & \mathrm{Age} (\mathrm{H}\delta_{\mathrm{F}}) & \mathrm{Age} (\mathrm{H}\delta_{\mathrm{A}}) & \mathrm{Age} (\mathrm{H}\gamma_{\mathrm{F}}) & \mathrm{<Age>} &\log(\mathrm{Age/yr}) & \mathrm{Age} \\
\hline

33 & 4.7^{+4.4}_{-2.3}      & \bf{5.8^{+9.2 }_{- 3.2}}  & 3.5^{+10.0}_{-2.5}         & \bf{5.4^{+8.8}_{-3.7}} & \bf{7.2^{+6.3}_{-3.6}}&5.32^{+7.74}_{-3.06} & 9.06^{+0.39}_{-1.14} & 1.14 ^{+1.67}_{-1.06}\\ 
34 &          -                    & \bf{9.9^{+0.8 }_{- 3.1}}  &    -                              &                -               &       -          &9.90^{+0.80}_{-3.1} & 9.10^{+0.91}_{-0.65} & 1.25 ^{+8.97}_{-0.97}\\ 
35 & 4.6^{+4.5}_{-2.9}      & 4.3^{+10.4}_{- 2.1}       &     -                             & 2.3^{+5.1}_{-0.9}        & 4.6^{+9.0}_{-2.8}&4.52^{+6.09}_{-1.76} & 6.91^{+1.43}_{-0.09} & 0.008^{+0.27}_{-0.02}\\ 
\hline		      	    		    	     	     	       	       	       	   	    	  	    
40 & 7.0^{+6.4}_{-2.9}       &                  -                & \bf{4.5^{+10.5}_{-2.8}} & \bf{6.9^{+8.1}_{-4.0}}  &         -        &6.13^{+8.30}_{-3.23} &-&-\\ 
42 & 5.9^{+9.1}_{-3.0}       & \bf{9.5^{+3.6 }_{- 4.2}}  & \bf{3.8^{+11.2}_{-2.5}}  & \bf{6.9^{+7.9}_{-3.9}}  &         -        &6.52^{+7.95}_{-3.4} &-&-\\
43 & 4.4^{+5.1}_{-2.5}       & \bf{5.6^{+8.5 }_{- 3.3}}  & 1.4^{+3.2 }_{-0.4}         & 1.7^{+3.7}_{-0.4}        & \bf{5.5^{+8.4}_{-3.4}}&3.72^{+5.78}_{-2} &-&-\\ 
45 & \bf{1.6^{+7.9}_{-0.6}} & \bf{2.7^{+10.8}_{- 1.7}}  &        -                          &                   -            & \bf{3.4^{+8.5}_{-2.4}}&2.56^{+9.06}_{-1.57} &-&-\\ 
47 & 6.9^{+8.1}_{-5.2}       & \bf{2.5^{+11.3}_{- 1.3}}  &                       -           & \bf{2.6^{+7.4}_{-1.3}} & \bf{2.0^{+11 }_{-1.0}}&3.50^{+9.45}_{-2.2} &-&-\\ 
\noalign{\smallskip}
     \hline
   \end{array}
   $$
Column (1): Globular cluster ID. Columns (2)  to (6): Derived ages. The index used
for its derivation is shown between brackets. Column (7): Average age. Columns (8) and (9): Ages
derived from GALEV.

\end{table}

In Table \ref{Edades} we show individual ages derived from the age diagnostic plots for each Balmer index, and the mean value considering all Balmer indices.
We do not extrapolated values outside the model grids and errors were calculated  in the same way as we did in Table \ref{Metal}. Values in boldface indicate
that the errors lie outside the model grid. The last two columns show the 
derived ages from GALEV (considering Z=0.004).  Ages derived from photometry 
do not agree with the values derived using spectroscopy. Photometric ages were
underestimated, compared with the spectroscopic ones.  This underestimation 
is more dramatic for the globular cluster ID=35 where the photometric age 
do not share the same order of magnitude as the spectroscopic ones. 
This result is not entirely unexpected, since old globular clusters are faint in
the U-band, which provides much of the leverage for age determinations.
Furthermore, model uncertainties and degeneracies in age/metallicity/reddening all 
combine to produce larger uncertainties on the photometric ages for old globular clusters.

%%%%%%%%%%%%%%%%%%%%%%%%%%%%%%%%%%%%%%%%%%%%%%%%%%%%%%%%%%%%%%%
\section{Discussion and conclusion}
%%%%%%%%%%%%%%%%%%%%%%%%%%%%%%%%%%%%%%%%%%%%%%%%%%%%%%%%%%%%%%%

Although uncertainties on the age estimates for our sample of globular clusters in
NGC~45 remain large, we were able to constrain their metallicities and
$\alpha$/Fe abundance ratios.
These showed that the globular clusters in NGC~45 are metal poor, corroborating the 
metal poor population deducted from the globular cluster colors in \citet{Mora}. 

Assuming  that  the globular clusters are tracer of star formation events, 
considering that NGC~45 is an isolated galaxy, probably a background object
and, not a true member of the Sculptor group \citep{puche}, it is puzzling 
how these entities formed in this galaxy. \citet{alan} suggested that NGC~45 
once made a close pass by the  Sculptor group, getting close to NGC~7793  
and transferring angular momentum. This would have been excited the 
globular cluster formation in NGC~45 but there is no  record of this 
in the globular cluster population. Therefore, the formation of the 
globular clusters  must be happen  at early times probably when the galaxy assembled. 

The globular cluster velocities and the velocity dispersion of the system argue in favor of a real  association 
between the globular clusters and the galaxy bulge. The velocity distribution 
of the GC population seems to be dominated by random motions, although 
it would be desirable to corroborate this statement with further 
velocity measurements of the remaining 11 candidates.

The sub-solar [$\alpha$/Fe] values derived here are unlike those
typically observed in old GC populations \citep[e.g.][]{puzia2005},
including those in the Milky Way. However, we note that they are
consistent with those derived by \citet{knut}. It is also of interest
to note that dwarf galaxies in the Local Group tend to show
less alpha-enhanced abundance ratios than the Milky Way \citep{Sbordone,Tolstoy2003}, 
despite the fact that many of them are satellite galaxies (e.g. LMC, SMC, etc).
This may point to important differences in the early chemical
evolution in these different types of galaxies, and is potentially
an argument against the notion that a major fraction of the GCs
in large galaxies could have been accreted from minor galaxies
similar to those observed today.

One possible explanation for the relatively low [$\alpha$/Fe] ratios
of the GCs in NGC~45 is that the formation and assembly of the
halo/bulge component took longer than in major galaxies like the
Milky Way. This would allow time for Type Ia supernovae to appear
and enrich the gas with greater amounts of Fe.

%%%%%%%%%%%%%%%%%%%%%%%%%%%%%%%%%%%%%%%%%%%%%%%%%%%%%%%%%%%%%%%
\begin{acknowledgements}
%%%%%%%%%%%%%%%%%%%%%%%%%%%%%%%%%%%%%%%%%%%%%%%%%%%%%%%%%%%%%%%

We would like to thank Thomas Puzia for providing us the GONZO code, Steffen Mieske for providing us the
elliptical galaxy spectra template, Laurent Chemin for gently provide us the NGC~45 H~I velocity field data, 
and the referee for his/her work.

\end{acknowledgements}

\bibliographystyle{aa}
\bibliography{Marcelo}

\section{Apendix}

\begin{landscape}
\begin{table}
\caption[Measured indice.]{Measured indice.
}
\label{calibrationT}
\resizebox{1\columnwidth}{!}{ 

 \begin{tabular}{rcccccccccc}           
   \hline              
   \hline              
 {Indice} $\backslash$ {ID}                                 &          &  33                                        &  34                                          & 35                                          &  40  				       &    42                                             &  43  				              &  45  				             &  47  				          \\     	
   \hline                                     				                                            						 						  											       							
 CN$_1$       $\pm \sigma ^{+\sigma_{vel}}_{-\sigma_{vel}}$ &     mag  &  $    -0.021 \pm 0.005^{+0.046}_{-0.048} $ & $ -0.020 \pm 0.008^{+0.009}_{-0.003} $       & $  ~~~0.012 \pm 0.005^{+0.048}_{-0.050} $   & $ -0.017 \pm 0.004^{+0.020}_{-0.027} $      &    $ -0.079 \pm 0.007 ^{+0.013}_{-0.013}  $       & $ -0.155 \pm 0.014^{+0.018}_{-0.001} $         & $ -0.066 \pm 0.017^{+0.001}_{-0.001} $       & $  ~~~0.007 \pm 0.029^{+0.052}_{-0.021} $  \\    
 CN$_2$       $\pm \sigma ^{+\sigma_{vel}}_{-\sigma_{vel}}$ &     mag  &  $    -0.038 \pm 0.003^{+0.004}_{-0.001} $ & $  ~~~0.012 \pm 0.005^{+0.000}_{-0.001} $    & $  ~~~0.001 \pm 0.004^{+0.000}_{-0.000} $   & $  ~~~0.004 \pm 0.003^{+0.003}_{-0.002} $   &    $ -0.052 \pm 0.004 ^{+0.000} _{-0.001} $       & $ -0.113 \pm 0.008^{+0.001}_{-0.001} $         & $  ~~~0.011 \pm 0.012^{+0.002}_{-0.022} $    & $  ~~~0.066 \pm 0.021^{+0.050}_{-0.030} $  \\    
 Ca4227       $\pm \sigma ^{+\sigma_{vel}}_{-\sigma_{vel}}$ &     \AA  &  $  ~~~0.705 \pm 0.135^{+0.073}_{-0.000} $ & $  ~~~0.905 \pm 0.177^{+0.002}_{-0.005} $    & $  ~~~0.783 \pm 0.173^{+0.017}_{-0.020} $   & $  ~~~0.047 \pm 0.104^{+0.014}_{-0.042} $   &    $  ~~~0.601 \pm 0.178 ^{+0.002} _{-0.008} $    & $  ~~~0.775 \pm 0.331^{+0.012}_{-0.020} $      & $  ~~~1.636 \pm 0.430^{+0.022}_{-0.022} $    & $  ~~~0.516 \pm 0.716^{+0.501}_{-0.336} $  \\    
 G4300        $\pm \sigma ^{+\sigma_{vel}}_{-\sigma_{vel}}$ &     \AA  &  $  ~~~4.632 \pm 0.195^{+0.854}_{-0.075} $ & $  ~~~5.399 \pm 0.214^{+1.144}_{-0.022} $    & $  ~~~4.876 \pm 0.257^{+0.173}_{-0.185} $   & $  ~~~2.167 \pm 0.134^{+0.075}_{-0.849} $   &    $  ~~~4.452 \pm 0.220 ^{+0.020} _{-0.051} $    & $  ~~~3.863 \pm 0.338^{+0.030}_{-0.025} $      & $  ~~~6.603 \pm 0.500^{+0.117}_{-0.113} $    & $  ~~~9.475 \pm 0.792^{+0.333}_{-0.063} $  \\    
 Fe4383       $\pm \sigma ^{+\sigma_{vel}}_{-\sigma_{vel}}$ &     \AA  &  $  ~~~1.579 \pm 0.238^{+0.009}_{-0.043} $ & $  ~~~2.518 \pm 0.237^{+0.011}_{-0.041} $    & $  ~~~0.951 \pm 0.303^{+0.073}_{-0.108} $   & $ -0.462 \pm 0.174^{+0.399}_{-0.068} $      &    $  ~~~2.270 \pm 0.248 ^{+0.029} _{-0.060} $    & $  ~~~0.384 \pm 0.366^{+0.089}_{-0.102} $      & $    -1.343 \pm 0.818^{+0.289}_{-0.119} $    & $ -0.944 \pm 0.889^{+0.538}_{-2.523} $     \\    
 Ca4455       $\pm \sigma ^{+\sigma_{vel}}_{-\sigma_{vel}}$ &     \AA  &  $  ~~~0.889 \pm 0.246^{+0.045}_{-0.015} $ & $  ~~~0.412 \pm 0.242^{+0.015}_{-0.081} $    & $  ~~~1.021 \pm 0.313^{+0.061}_{-0.027} $   & $  ~~~0.260 \pm 0.180^{+0.025}_{-0.023} $   &    $  ~~~0.666 \pm 0.253 ^{+0.046} _{-0.011} $    & $  ~~~0.895 \pm 0.370^{+0.048}_{-0.009} $      & $  ~~~0.158 \pm 0.844^{+0.040}_{-0.090} $    & $ -0.935 \pm 0.950^{+0.896}_{-0.260} $     \\    
 Fe4531       $\pm \sigma ^{+\sigma_{vel}}_{-\sigma_{vel}}$ & 	  \AA  &  $  ~~~2.112 \pm 0.258^{+0.056}_{-0.037} $ & $  ~~~2.847 \pm 0.257^{+0.025}_{-0.018} $    & $  ~~~1.342 \pm 0.323^{+0.005}_{-0.042} $   & $  ~~~0.174 \pm 0.197^{+0.014}_{-0.006} $   &    $  ~~~2.438 \pm 0.266 ^{+0.051} _{-0.043} $    & $  ~~~2.881 \pm 0.391^{+0.122}_{-0.055} $      & $  ~~~1.085 \pm 0.873^{+0.007}_{-0.441} $    & $  ~~~3.993 \pm 1.026^{+0.875}_{-0.468} $  \\    
 Fe4668       $\pm \sigma ^{+\sigma_{vel}}_{-\sigma_{vel}}$ & 	  \AA  &  $  ~~~1.640 \pm 0.358^{+0.042}_{-0.055} $ & $  ~~~3.526 \pm 0.285^{+0.001}_{-0.026} $    & $ -1.152 \pm 0.404^{+0.072}_{-0.032} $      & $ -3.567 \pm 0.301^{+0.045}_{-0.225} $      &    $  ~~~1.541 \pm 0.301 ^{+0.037} _{-0.657} $    & $  ~~~0.464 \pm 0.431^{+0.018}_{-0.015} $      & $ -2.511 \pm 0.921^{+0.226}_{-1.833} $       & $  ~~~4.999 \pm 1.101^{+2.516}_{-1.288} $  \\    
 H$_\beta$    $\pm \sigma ^{+\sigma_{vel}}_{-\sigma_{vel}}$ &     \AA  &  $  ~~~3.045 \pm 0.360^{+0.003}_{-0.000} $ & $  ~~~1.287 \pm 0.287^{+0.016}_{-0.021} $    & $  ~~~2.810 \pm 0.407^{+0.019}_{-0.446} $   & $  ~~~2.995 \pm 0.302^{+0.009}_{-0.000} $   &    $  ~~~2.707 \pm 0.304 ^{+0.004} _{-0.004} $    & $  ~~~2.743 \pm 0.433^{+0.005}_{-0.002} $      & $  ~~~3.233 \pm 0.928^{+0.028}_{-0.087} $    & $  ~~~2.230 \pm 1.115^{+0.100}_{-0.145} $  \\    
 Fe5015       $\pm \sigma ^{+\sigma_{vel}}_{-\sigma_{vel}}$ & 	  \AA  &  $  ~~~2.108 \pm 0.376^{+0.091}_{-0.046} $ & $  ~~~1.136 \pm 0.328^{+0.054}_{-0.034} $    & $  ~~~1.740 \pm 0.422^{+0.155}_{-0.977} $   & $  ~~~1.915 \pm 0.318^{+0.123}_{-0.048} $   &    $  ~~~2.563 \pm 0.336 ^{+0.093} _{-0.009} $    & $  ~~~2.716 \pm 0.442^{+0.214}_{-0.411} $      & $  ~~~5.323 \pm 0.954^{+0.200}_{-0.042} $    & $  ~~~0.887 \pm 1.157^{+0.761}_{-0.157} $  \\    
 Mg$_1$       $\pm \sigma ^{+\sigma_{vel}}_{-\sigma_{vel}}$ &     mag  &  $  ~~~0.032 \pm 0.010^{+0.000}_{-0.000} $ & $  ~~~0.030 \pm 0.010^{+0.000}_{-0.000} $    & $  ~~~0.004 \pm 0.012^{+0.000}_{-0.009} $   & $ -0.041 \pm 0.007^{+0.000}_{-0.000} $      &    $    -0.006 \pm 0.010 ^{+0.000} _{-0.000} $    & $  ~~~0.075 \pm 0.011^{+0.000}_{-0.000} $      & $  ~~~0.066 \pm 0.029^{+0.026}_{-0.011} $    & $ -0.024 \pm 0.040^{+0.028}_{-0.025} $     \\    
 Mg$_2$       $\pm \sigma ^{+\sigma_{vel}}_{-\sigma_{vel}}$ & 	  mag  &  $  ~~~0.090 \pm 0.010^{+0.000}_{-0.000} $ & $  ~~~0.107 \pm 0.010^{+0.000}_{-0.000} $    & $  ~~~0.092 \pm 0.012^{+0.000}_{-0.006} $   & $  ~~~0.050 \pm 0.007^{+0.000}_{-0.000} $   &    $  ~~~0.074 \pm 0.010 ^{+0.000} _{-0.000} $    & $  ~~~0.118 \pm 0.011^{+0.000}_{-0.000} $      & $  ~~~0.124 \pm 0.029^{+0.032}_{-0.014} $    & $  ~~~0.104 \pm 0.040^{+0.042}_{-0.033} $  \\    
 Mg$_b$       $\pm \sigma ^{+\sigma_{vel}}_{-\sigma_{vel}}$ & 	  \AA  &  $  ~~~0.793 \pm 0.398^{+0.042}_{-0.048} $ & $  ~~~1.389 \pm 0.337^{+0.013}_{-0.020} $    & $  ~~~1.270 \pm 0.444^{+0.004}_{-0.003} $   & $  ~~~0.705 \pm 0.333^{+0.036}_{-0.002} $   &    $  ~~~1.235 \pm  0.345^{+0.014} _{-0.023} $    & $  ~~~1.120 \pm 0.447^{+0.041}_{-0.026} $      & $  ~~~2.111 \pm 0.984^{+0.095}_{-0.087} $    & $  ~~~1.586 \pm 1.233^{+0.198}_{-0.148} $  \\    
 Fe5270       $\pm \sigma ^{+\sigma_{vel}}_{-\sigma_{vel}}$ & 	  \AA  &  $  ~~~1.315 \pm 0.402^{+0.036}_{-0.021} $ & $  ~~~1.358 \pm 0.347^{+0.024}_{-0.050} $    & $  ~~~1.338 \pm 0.450^{+0.055}_{-0.053} $   & $  ~~~0.574 \pm 0.336^{+0.073}_{-0.052} $   &    $  ~~~1.062 \pm 0.355 ^{+0.038} _{-0.073} $    & $  ~~~1.764 \pm 0.451^{+0.114}_{-0.593} $      & $  ~~~1.240 \pm 0.993^{+0.047}_{-0.670} $    & $  ~~~1.968 \pm 1.239^{+0.111}_{-0.156} $  \\    
 Fe5335       $\pm \sigma ^{+\sigma_{vel}}_{-\sigma_{vel}}$ & 	  \AA  &  $  ~~~0.737 \pm 0.417^{+0.043}_{-0.042} $ & $  ~~~1.470 \pm 0.350^{+0.042}_{-0.040} $    & $  ~~~0.952 \pm 0.464^{+0.054}_{-0.052} $   & $  ~~~0.439 \pm 0.350^{+0.029}_{-0.040} $   &    $  ~~~0.823 \pm  0.368^{+0.378} _{-0.512} $    & $  ~~~1.272 \pm 0.454^{+0.013}_{-0.036} $      & $  ~~~1.483 \pm 1.001^{+0.045}_{-0.061} $    & $  ~~~0.990 \pm 1.250^{+0.332}_{-0.206} $  \\    
 Fe5406       $\pm \sigma ^{+\sigma_{vel}}_{-\sigma_{vel}}$ &     \AA  &  $  ~~~0.808 \pm 0.419^{+0.004}_{-0.013} $ & $  ~~~0.894 \pm 0.354^{+0.008}_{-0.010} $    & $  ~~~0.655 \pm 0.467^{+0.019}_{-0.013} $   & $  ~~~0.402 \pm 0.352^{+0.014}_{-0.010} $   &    $  ~~~0.965 \pm  0.371^{+0.028} _{-0.010} $    & $  ~~~1.123 \pm 0.456^{+0.009}_{-0.337} $      & $  ~~~1.169 \pm 1.012^{+0.062}_{-0.163} $    & $  ~~~1.941 \pm 1.258^{+1.184}_{-0.550} $  \\    
 Fe5709       $\pm \sigma ^{+\sigma_{vel}}_{-\sigma_{vel}}$ & 	  \AA  &  $  ~~~0.380 \pm 0.419^{+0.004}_{-0.001} $ & $  ~~~0.770 \pm 0.356^{+0.001}_{-0.021} $    & $  ~~~0.704 \pm 0.468^{+0.003}_{-0.019} $   & $  ~~~0.172 \pm 0.353^{+0.008}_{-0.004} $   &    $  ~~~0.742 \pm  0.372^{+0.014} _{-0.005} $    & $  ~~~0.674 \pm 0.457^{+0.033}_{-0.028} $      & $  ~~~0.598 \pm 1.032^{+0.073}_{-0.002} $    & $  ~~~1.910 \pm 1.270^{+1.108}_{-0.476} $  \\    
% Fe5782       $\pm \sigma ^{+\sigma_{vel}}_{-\sigma_{vel}}$ & 	  \AA  &  $  ~~~0.247 \pm 0.421^{+0.039}_{-0.000} $ & $  ~~~0.427 \pm 0.359^{+0.004}_{-0.016} $    & $  ~~~0.223 \pm 0.471^{+0.045}_{-0.360} $   & $ -0.026 \pm 0.355^{+0.019}_{-0.000} $      &    $  ~~~0.222 \pm  0.376^{+0.043} _{-0.015} $    & $  ~~~0.614 \pm 0.462^{+0.026}_{-0.053} $      & $ -0.944 \pm 1.049^{+0.094}_{-0.100} $       & $ -0.391 \pm 1.282^{+0.125}_{-0.247} $     \\    
% Na$_D$       $\pm \sigma ^{+\sigma_{vel}}_{-\sigma_{vel}}$ & 	  \AA  &  $  ~~~0.656 \pm 0.428^{+0.044}_{-0.021} $ & $  ~~~0.311 \pm 0.376^{+0.214}_{-0.012} $    & $  ~~~0.638 \pm 0.479^{+0.255}_{-0.060} $   & $  ~~~0.139 \pm 0.362^{+0.052}_{-0.002} $   &    $  ~~~0.878 \pm 0.394 ^{+0.020} _{-0.034} $    & $  ~~~0.725 \pm 0.467^{+0.037}_{-0.066} $      & $ -1.389 \pm 1.061^{+0.263}_{-0.512} $       & $ -2.383 \pm 1.300^{+0.188}_{-1.443} $     \\    
 H$\delta_A$  $\pm \sigma ^{+\sigma_{vel}}_{-\sigma_{vel}}$ &     \AA  &  $  ~~~3.696 \pm 0.435^{+0.038}_{-0.000} $ & $  ~~~0.388 \pm 0.387^{+0.053}_{-0.003} $    & $  ~~~3.686 \pm 0.490^{+0.008}_{-0.018} $   & $  ~~~4.621 \pm 0.367^{+0.032}_{-0.028} $   &    $  ~~~2.698 \pm  0.405^{+0.059} _{-0.006} $    & $  ~~~3.850 \pm 0.477^{+0.077}_{-0.036} $      & $ -1.843 \pm 1.330^{+1.655}_{-1.064} $       & $  ~~~1.876 \pm 1.361^{+0.105}_{-4.721} $  \\    
 H$\gamma_A$  $\pm \sigma ^{+\sigma_{vel}}_{-\sigma_{vel}}$ &     \AA  &  $  ~~~1.915 \pm 0.468^{+0.092}_{-0.170} $ & $ -1.337 \pm 0.396^{+0.051}_{-0.066} $       & $  ~~~1.146 \pm 0.515^{+0.136}_{-0.172} $   & $  ~~~1.895 \pm 0.394^{+0.462}_{-0.040} $   &    $  ~~~0.014 \pm 0.413 ^{+0.056} _{-0.049} $    & $  ~~~0.220 \pm 0.489^{+0.203}_{-0.073} $      & $ -0.069 \pm 1.345^{+0.129}_{-1.716} $       & $  ~~~0.234 \pm 1.397^{+1.350}_{-0.674} $  \\    
 H$\delta_F$  $\pm \sigma ^{+\sigma_{vel}}_{-\sigma_{vel}}$ &     \AA  &  $  ~~~2.972 \pm 0.478^{+0.003}_{-0.008} $ & $  ~~~0.997 \pm 0.399^{+0.022}_{-0.011} $    & $  ~~~3.716 \pm 0.527^{+0.024}_{-0.027} $   & $  ~~~3.462 \pm 0.399^{+0.026}_{-0.035} $   &    $  ~~~2.560 \pm 0.418 ^{+0.019} _{-0.451} $    & $  ~~~2.962 \pm 0.494^{+0.019}_{-0.000} $      & $  ~~~0.135 \pm 1.372^{+0.508}_{-0.250} $    & $  ~~~0.266 \pm 1.443^{+0.239}_{-0.018} $  \\    
 H$\gamma_F$  $\pm \sigma ^{+\sigma_{vel}}_{-\sigma_{vel}}$ &     \AA  &  $  ~~~2.372 \pm 0.481^{+0.046}_{-0.342} $ & $  ~~~0.678 \pm 0.402^{+0.003}_{-0.008} $    & $  ~~~2.263 \pm 0.539^{+0.060}_{-0.081} $   & $  ~~~2.180 \pm 0.403^{+0.134}_{-0.146} $   &    $  ~~~1.400 \pm 0.421 ^{+0.000} _{-0.008} $    & $  ~~~1.786 \pm 0.496^{+0.000}_{-0.020} $      & $  ~~~1.456 \pm 1.378^{+0.090}_{-0.081} $    & $  ~~~2.060 \pm 1.462^{+0.283}_{-0.105} $  \\

\noalign{\smallskip}
\hline
\end{tabular}
}   
 $\sigma$ corresponds to the error from the random noise. $\sigma_{vel}$ corresponds to
the error due to the velocity uncertainties.

\end{table}

\end{landscape}

\end{document}